\documentclass[prl,aps,floats,twocolumn,showpacs]{revtex4}

\usepackage{graphicx}
\usepackage{amssymb} 
\usepackage{amsmath}

\begin{document}

\def\gtrsim{\mathrel{\hbox{\rlap{\hbox{\lower4pt\hbox{$\sim$}}}\hbox{$>$}}}}

\title{Observational Consequences of Dark Energy Decay}
\author{
Ue-Li  Pen$^{1}$\thanks{E-mail:\ pen@cita.utoronto.ca} and
Pengjie Zhang$^2$\thanks{E-mail:\ pjzhang@shao.ac.cn} 
}
\date{\today}

\begin{abstract}

We consider the generic scenario of dark energy which arises through
the latent heat of a hidden sector first order cosmological phase
transition.  This field could account for the extra radiation degree
of freedom suggested by the CMB. We present the bubble nucleation
solution for the viscous limit.  The decay rate of the field is
constrained by published KSZ data, and may be an explanation of
current excess ISW correlations.  Cross correlation of current and
future surveys can further constrain or test the parameter space.  The
decay model is plausibly in the observable range, and avoids anthropic
problems.  This class of models is not well constrained by the popular
dark energy figure of merit.

\end{abstract}
\maketitle

\newcommand{\be}{\begin{eqnarray}}
\newcommand{\ee}{\end{eqnarray}}
\newcommand{\beq}{\begin{equation}}
\newcommand{\eeq}{\end{equation}}

{\it Introduction.} -- Dark Energy is one of the most mysterious
puzzles in modern physics.  In this paper we consider the possibility
that dark energy is a mundane first order phase transition in a hidden
sector thermal field.  Such phase transitions have been proposed at
these low temperatures\cite{Haim2000153,2009PhRvD..79j3504D}, and the
energy scales might arise naturally in a seesaw
mechanism\cite{2004JCAP...10..011C}. Standard cosmological phase
transitions are described in textbooks\cite{1990eaun.book.....K}. In
order for a false vacuum to sustain an accelerating epoch dominated by
dark energy, the phase transition must be strongly first order. The
nucleation to the true vacuum can occur through quantum and thermal
tunnelling processes.  The rates depend exponentially on the details
of the potential.  In order for the universe to appear accelerating
over a substantial history, the rates cannot be much higher than the
inverse age of the universe.  Thus, the phase transition must occur
through the nucleation of discrete bubbles.  If the false vacuum life
time is shorter than 19 Gyr\cite{2008PhRvD..78f3535P}, the phase
transition will eventually complete through the collision of bubbles,
which have sizes and separations comparable to the visible universe
today. Otherwise, the expansion rate of the universe wins over the
nucleation rate, and the vacuum dominated regions dominate in volume.
This has been interpreted to lead to problems with Boltzman brains
(ibid).  So it is interesting to observationally test if the
nucleation rate might indeed be high enough to eliminate this
potential problem.

It has been debated how generic a very strongly first order phase
transition is\cite{PhysRevD.77.063519}, and some amount of tuning is
required to be ``just right'' for it to explain the supernova data.
We feel that this amount of tuning is modest compared to that required
for a slow roll scenario, which is the most fashionable dark energy
model today\cite{2006astro.ph..9591A}.

We compute the physical properties of nucleating bubbles, and
observational consequences if we live inside one.  The bubble boundary
expands at close to the speed of light, and the line of sight will
generally not intersect a bubble boundary unless we live inside it. We
find that most of the dark energy decays into a fluid concentrated in
the outer 0.01\% of the bubble radius.  KSZ constrains our region to
have nucleated not more than about 3 Gyr ago. We propose future
observational tests for these bubbles, which include KSZ cross
correlations, and high precision dipole anisotropy searches in the
local matter distribution.

{\it Scenario.} -- The simplest scenario is a hidden sector radiation
fluid which currently only interacts with baryonic matter through
gravity, much like dark matter.  Current CMB data suggests that an
extra radiative degree of freedom may be
present\cite{2011ApJS..192...18K}, at a temperature comparable to that
of neutrinos.  The number of radiation degrees of freedom is usually
quoted as the effective number of neutrino species, $N_{\rm
  eff}=4.34^{+0.86}_{-0.88}$.  There are 3 known neutrino species,
leaving room for an extra light field or radiation fluid at $z\sim
1000$.  As the universe expands and cools, this fluid could undergo a
first order phase transition.  For a phase transition temperature of
$\sim 20$K, at $z\sim 10$, the resulting false vacuum energy could
explain today's apparent cosmological constant. As the universe
expands, the false vacuum state can tunnel to the true vacuum, either
thermally or quantum mechanically.  If the tunnelling rate is
comparable to today's universe age, the universe could exit the
trapped dark energy state through bubble nucleation.  During the
factor of ten cosmic expansion since the false vacuum trapping, the
fluid has cooled to a tenth of the potential energy difference.  The
decaying vacuum will expand into this fluid.  If viscosity is
neglible, the vacuum energy decay will all end up in the kinetic
energy of the bubble wall, which would then move at ultra relativistic
speeds. In the presence of viscosity, the wall sweeps up the ambient
radiation fluid, and the vacuum decay generates entropy in the fluid,
which moves outward.  The details of the bubble wall not important,
all the energetics is dominated by the swept up fluid and the entropy
generated by the decayed vacuum.  We shall solve the dynamics in the
next section.

\begin{figure}
\begin{center}
    \includegraphics[width=3.1in]{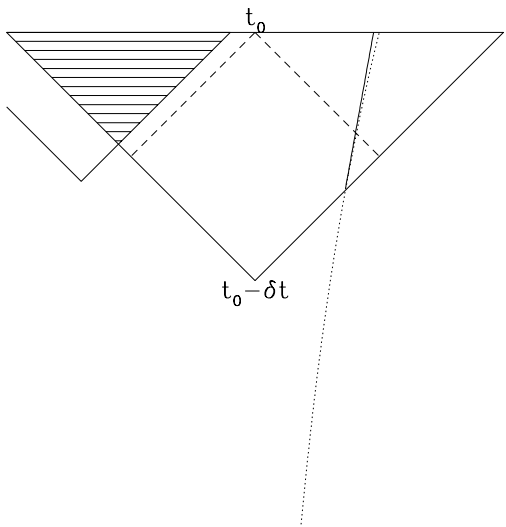}
  \end{center}
  \caption{bubble nucleation schematic.  $t_0$ labels a present day
    observer centered at a bubble nucleation event which occurred
    $\delta t=r_0/c$ in the past.  The last triangle denotes the true vacuum
  region. The dotted line denotes the trajectory of a dark matter
  shell in a $\Lambda$CDM cosmology.  The solid tangent to this line
  denotes the actual post nucleation trajectory.  The deviation of
  these two lines is observed on the past light cone indicated by the
  dashed line.  The parameters correspond to a bubble nucleation event
  half a Hubble time in the past.  On the left is a schematic
  second (unobservable) bubble nucleation event which occurs outside
  our past light cone.  The shaded region is the collision region
  of the bubbles, inside of which our analysis is not applicable.}  
\label{fig:cone}
\end{figure}

\begin{figure}
\begin{center}
    \includegraphics[width=3.1in]{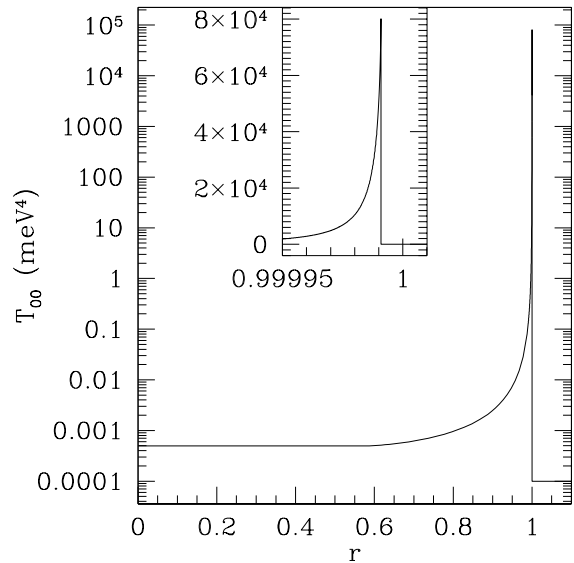}
  \end{center}
  \caption{bubble density profile.  Plotted is the dark matter frame
    00 component of the 
  radiation fluid stress energy tensor $\rho \gamma^2$, which is the
  conserved matter 
  quantity.  The inset shows the shock front on a linear scale.} 
\label{fig:density}
\end{figure}

{\it Dynamics.} --
The hydrodynamic solution of decay of vacuum energy was solved in
\cite{1998ApJ...509..537P}.  In the cosmological context, we are in
the limit of a relativistic perfect ambient fluid and a positive
potential energy, with about 10,000 times the density of the fluid.
In a bubble nucleation, this potential energy is released and results
in raising the entropy of the ambient fluid, and accelerating the
fluid outward, at a Lorenz factor $\gamma \sim 100$.  The numerical
solution is shown in figure \ref{fig:density}.  We note that actual
phase transitions can be more complex depending on the properties of
the viscosity\cite{2010JCAP...06..028E}.  Because most of the energy
always piles up near the bubble wall, the infinite viscosity limit and
zero viscosity limits result in similar net observables.

Most of the converted energy piles up within about $1/\gamma^2$ of the
bubble wall, with a local overdensity of $\sim \gamma^2$.  As we will
see below, a bubble might be as big as $\sim$ Gpc, which would have a
thickness of a few kpc.  A bubble edge passing through our galaxy or solar
system would have minimal impact, except on the most weakly bound
structures.  The Oort cloud could experience non-neglible
perturbations, perhaps resulting in enhanced cometary impacts on
earth.  The dynamical time at the Oort cloud is $\sim 10^7$ yr.  At
this radius, the dark energy contributes a $\sim 10^{-3}$ fractional
force of gravity.  A bubble passage changes this force abruptly (shock
impulse), potentially repopulating the loss cone that enters the inner
solar system. While small compared to galactic tides, it is also far
out of the adiabatic limit, with a correspondingly larger effect.  It
would be tempting to associate a bubble passage with the dinosaur
extinction event 65 Myr ago.

The dynamics of dark matter and baryons inside a bubble is
straightforward on subhorizon scales.  Each expanding matter shell
suddenly stops its dark energy acceleration when the shell passes, and
continues moving on a roughly straight trajectory.  For an observer
near the center of the nucleation event the universe still appears
isotropic, but not homogeneous.

We first compute the test particle evolution centered on the
nucleation event.  We make the simplifying assumptions that we
nucleate from a vacuum dominated phase, and neglect the finite
contribution of dark matter.  These assumptions are good to 20\%
today, and are progressively worse in the past.
We denote the present time by $t_0$.  We work in
the limit where the Hubble $H_0$ constant is the cosmological constant
$\Lambda=H_0^2$. The bubble nucleates at a lookback time $\delta t=r_0/c \ll
1/H_0$.  Each shell labelled by $m=cz/H_0$ deviates from its accelerated
expansion rate by a (inward) peculiar velocity
\be
\label{eqn:vp}
v_p = \frac{H_0^2 m(r_0-2m)}{c}
\ee
The peak blueshift is reached halfway at $m=r_0/4$.  The peak peculiar
velocity relative to the unperturbed (CMB) frame
is $v_{\rm max}=(H_0 r_0)^2/8c$.

{\it Statistics.} --
We can currently be living in the false vacuum, in which case
we would be unlikely to see any other bubbles on our past lightcone,
since any such bubble would expand at close to the speed of light.
The chance of perfect timing coincidence that we could see both walls
of a bubble are $\sim 10^{-4}$.  We can also live inside either one
bubble, or in the past history of more than one.

\begin{figure}
\begin{center}
    \includegraphics[width=3.1in]{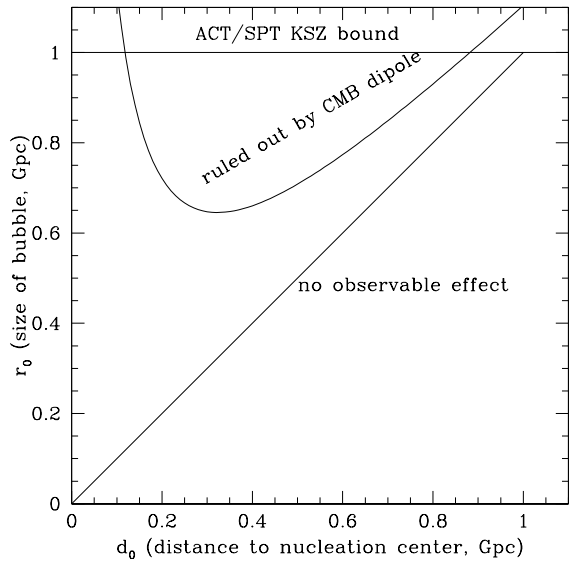}
  \end{center}
  \caption{
Bubble size and position allowed by CMB data.
}
\label{fig:rd}
\end{figure}

If we live inside a true vacuum bubble, we could either live near the
center, half way to the edge, or near the
edge\cite{2002APh....17..367A}.  We will go through each of these
possibilities.  Figure \ref{fig:rd} plots the parameter space that is
allowed by observations.

The CMB is an important constraint on isotropy and
homogeneity\cite{1995PhRvD..51.1525M}.  Locating us half way out to
the bubble edge maximizes the local velocity relative to the CMB rest
frame.  If this is larger than the observed CMB dipole, the parameter
space is ruled out.  Our observed dipole is 700 km/sec.  The local
peculiar velocity can cancel out some of the bulk flow.  We use a bulk
flow of 2000 km/sec for purposes of setting an upper bound.

The change of background cosmology inside the bubble leads to large
scale bulk flows.  One of the most sensitive probes is the kinetic
Sunyaev-Zeldovich (kSZ) effect\cite{2011PhRvL.107d1301Z}.  An observer at
the center of the bubble sees matter within the true vacuum bubble at
a systematic blueshift relative to the CMB: a matter overdensity
results in a higher CMB temperature, and thus a positive correlation
between matter and the CMB.

\begin{figure}
\begin{center}
    \includegraphics[width=3.1in]{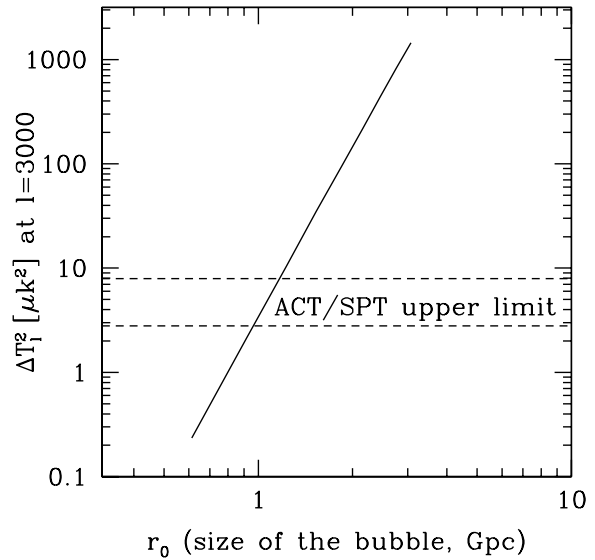}
  \end{center}
  \caption{The bubble generated kSZ effect, as seen by observers at the
    bubble center. The steep $r_0$ dependence largely comes from the peak
    bulk velocity dependence on $r_0$ ($v_{\rm max}\propto r_0^2$). The
    observed ACT/SPT upper limits constrain $r_0<1$ Gpc, even  if we neglect
    all other contributions to the kSZ effect. }
\label{fig:ksz}
\end{figure}
{\it The kinetic Sunyaev Zel'dovich effect.} --
The diffuse kSZ effect has not been detected yet. However, the upper limit of
the kSZ power spectrum is tightened significantly by ongoing 
experiments such as ACT and SPT. The latest upper limit of the kSZ power
spectrum $\Delta T_\ell^2$ at $\ell=3000$ is $2.8$-$6.5\mu$K$^2$ for SPT
\cite{2011ApJ...736...61S,2011arXiv1111.0932R}and $8\mu$K$^2$ for 
ACT\cite{2011ApJ...739...52D}. Following the same procedure as in
\cite{2011PhRvL.107d1301Z}, we 
calculate the bubble generated kSZ effect, as seen by an observer at the center
(Fig. \ref{fig:ksz}).  The kSZ power spectrum increases steeply with
increasing $r_0$ ($\Delta T_\ell^2\propto r_0^{\sim 5.5}$ \footnote{This
  strong $r_0$ dependence can be roughly understood combining Eq. 5 in
  \cite{2011PhRvL.107d1301Z} and Eq. \ref{eqn:vp}. From Eq. \ref{eqn:vp}, the
  peak velocity $v_{\rm max}\propto r_0^2$. If we are able to neglect the redshift
  dependence of $\Delta^2_e$ and approximate $1+z\simeq z$ since $z\ll 1$, we
  obtain $\Delta T_\ell^2\propto r_0^{\sim 6}$.  From this argument, the
  strong $r_0$ dependence mainly comes from the peak velocity dependence on
  $r_0$. } ), making the kSZ
effect a sensitive test of dark energy decay. The kSZ test puts a constraint
$r_0<1$ Gpc (Fig. \ref{fig:ksz}). This constraint assumes no other kSZ
contributions. In reality, the kSZ contributions 
from free electrons after reionization(e.g. \cite{2004MNRAS.347.1224Z})  and
from patchy reionization can both 
reach a few $\mu$K$^2$ (e.g \cite{2011arXiv1111.6386Z}).  In this sense, the
kSZ constraint $r_0<1$ Gpc is 
rather conservative.

{\it Matter Anisotropy.} --
For an observer not centered at the nucleation event, the mean cosmic
density will appear larger in the direction of the nucleation center.
The observer's peculiar velocity vector relative to the CMB points
towards this center as well.  The Hubble constant exhibits
anisotropies. These effects all scale as $(H_0\delta t)^2$.  We have
seen above that the KSZ constraints are less than 1\%.  A direct
dipole search would need to make precise measurements of a local dipole.

{\it Cross Correlation.} -- More sensitive tests arise in a cross
correlation of the distribution of matter with the CMB.  The bubble
induced kSZ has the same sign as the ISW effect, which positively
correlates matter fluctuations with CMB temperature.  Since we have
already constrained the bubble flow peak to occur at $z\lesssim 0.05$,
a shallow all sky survey is most sensitive.  There are indications of
such positive detections\cite{2006MNRAS.372L..23C} in an SDSS-WMAP
cross correlation, which appear larger than expected for pure ISW.
Similarly, it produces a 2MASS-WMAP cross correlation at sub-degree
scales, which can be compared to the current weak reported
correlations\cite{2008PhRvD..78d3519H}.  A cross correlation of with
Planck could confirm this detection.  The peak velocity $v_{\rm
  max}\simeq 2\times 10^3 (r_0/{\rm Gpc})^2$ km$/s$ at redshift
$z_{\rm max} \simeq 0.07 (r_0/{\rm Gpc})^2 $. Since there are no
published 2MASS-Planck measurements, we will not perform numerical
calculation of the cross correlation signal here. We use a simple
scaling.  The dark flow generated kSZ cross correlation
(Fig.~\ref{fig:ksz}, \cite{2010MNRAS.407L..36Z}) predicts a $\sim
10\sigma$ detection of a bubble with $r_0=1$ Gpc through 2MASS-Planck
cross correlation.

{\it Conclusion.} --
We have studied the physical properties and observational consequences
of recent bubble nucleation events.  These events are generic
consequences if our dark energy is a false vacuum due to a strongly
first order hidden sector phase transition, and depends only weakly on
the microscopic properties of the field.  They may account for the
current enhanced SDSS-WMAP correlations.  Further studies with Planck
and 2MASS have the sensitivity to confirm the presence of these
bubbles.  Previous predictions of decay rates shorter than $\sim 20$
Gyr likely result in observable consequences.

We thank Eric Switzer and Eugene Lim and for helpful discussions. ULP
thanks NSERC and SHAO for financial support. PJZ Thanks the support by
the NSFC grants and the Beyond the Horizons program.

\newcommand{\araa}{ARA\&A}   
\newcommand{\afz}{Afz}       
\newcommand{\aj}{AJ}         
\newcommand{\azh}{AZh}       
\newcommand{\aaa}{A\&A}      
\newcommand{\aas}{A\&AS}     
\newcommand{\aar}{A\&AR}     
\newcommand{\apjs}{ApJS}     
\newcommand{\apjl}{ApJ}      
\newcommand{\apss}{Ap\&SS}   
\newcommand{\baas}{BAAS}     
\newcommand{\jaa}{JA\&A}     
\newcommand\jcap{{J. Cosmology Astropart. Phys.}}%
\newcommand{\mnras}{MNRAS}   
\newcommand{\pasj}{PASJ}     
\newcommand{\pasp}{PASP}     
\newcommand{\paspc}{PASPC}   
\newcommand{\qjras}{QJRAS}   
\newcommand{\sci}{Sci}       
\newcommand{\sova}{SvA}      
\newcommand{\aap}{A\&A}

\bibliography{bubble}

\begin{thebibliography}{21}
\expandafter\ifx\csname natexlab\endcsname\relax\def\natexlab#1{#1}\fi
\expandafter\ifx\csname bibnamefont\endcsname\relax
  \def\bibnamefont#1{#1}\fi
\expandafter\ifx\csname bibfnamefont\endcsname\relax
  \def\bibfnamefont#1{#1}\fi
\expandafter\ifx\csname citenamefont\endcsname\relax
  \def\citenamefont#1{#1}\fi
\expandafter\ifx\csname url\endcsname\relax
  \def\url#1{\texttt{#1}}\fi
\expandafter\ifx\csname urlprefix\endcsname\relax\def\urlprefix{URL }\fi
\providecommand{\bibinfo}[2]{#2}
\providecommand{\eprint}[2][]{\url{#2}}

\bibitem[{\citenamefont{{Page}}(2008)}]{2008PhRvD..78f3535P}
\bibinfo{author}{\bibfnamefont{D.~N.} \bibnamefont{{Page}}},
  \bibinfo{journal}{\prd} \textbf{\bibinfo{volume}{78}}, \bibinfo{eid}{063535}
  (\bibinfo{year}{2008}), \eprint{arXiv:hep-th/0610079}.

\bibitem[{\citenamefont{Goldberg}(2000)}]{Haim2000153}
\bibinfo{author}{\bibfnamefont{H.}~\bibnamefont{Goldberg}},
  \bibinfo{journal}{Physics Letters B} \textbf{\bibinfo{volume}{492}},
  \bibinfo{pages}{153 } (\bibinfo{year}{2000}), ISSN \bibinfo{issn}{0370-2693},
  \urlprefix\url{http://www.sciencedirect.com/science/article/pii/S0370269300010455}.

\bibitem[{\citenamefont{{Dutta} et~al.}(2009)\citenamefont{{Dutta}, {Hsu},
  {Reeb}, and {Scherrer}}}]{2009PhRvD..79j3504D}
\bibinfo{author}{\bibfnamefont{S.}~\bibnamefont{{Dutta}}},
  \bibinfo{author}{\bibfnamefont{S.~D.~H.} \bibnamefont{{Hsu}}},
  \bibinfo{author}{\bibfnamefont{D.}~\bibnamefont{{Reeb}}}, \bibnamefont{and}
  \bibinfo{author}{\bibfnamefont{R.~J.} \bibnamefont{{Scherrer}}},
  \bibinfo{journal}{\prd} \textbf{\bibinfo{volume}{79}}, \bibinfo{eid}{103504}
  (\bibinfo{year}{2009}), \eprint{0902.4699}.

\bibitem[{\citenamefont{{Chacko} et~al.}(2004)\citenamefont{{Chacko}, {Hall},
  and {Nomura}}}]{2004JCAP...10..011C}
\bibinfo{author}{\bibfnamefont{Z.}~\bibnamefont{{Chacko}}},
  \bibinfo{author}{\bibfnamefont{L.~J.} \bibnamefont{{Hall}}},
  \bibnamefont{and} \bibinfo{author}{\bibfnamefont{Y.}~\bibnamefont{{Nomura}}},
  \bibinfo{journal}{\jcap} \textbf{\bibinfo{volume}{10}}, \bibinfo{pages}{11}
  (\bibinfo{year}{2004}), \eprint{arXiv:astro-ph/0405596}.

\bibitem[{\citenamefont{{Kolb} and {Turner}}(1990)}]{1990eaun.book.....K}
\bibinfo{author}{\bibfnamefont{E.~W.} \bibnamefont{{Kolb}}} \bibnamefont{and}
  \bibinfo{author}{\bibfnamefont{M.~S.} \bibnamefont{{Turner}}},
  \emph{\bibinfo{title}{{The early universe.}}} (\bibinfo{year}{1990}).

\bibitem[{\citenamefont{M\'egevand and S\'anchez}(2008)}]{PhysRevD.77.063519}
\bibinfo{author}{\bibfnamefont{A.}~\bibnamefont{M\'egevand}} \bibnamefont{and}
  \bibinfo{author}{\bibfnamefont{A.~D.} \bibnamefont{S\'anchez}},
  \bibinfo{journal}{Phys. Rev. D} \textbf{\bibinfo{volume}{77}},
  \bibinfo{pages}{063519} (\bibinfo{year}{2008}),
  \urlprefix\url{http://link.aps.org/doi/10.1103/PhysRevD.77.063519}.

\bibitem[{\citenamefont{{Albrecht} et~al.}(2006)\citenamefont{{Albrecht},
  {Bernstein}, {Cahn}, {Freedman}, {Hewitt}, {Hu}, {Huth}, {Kamionkowski},
  {Kolb}, {Knox} et~al.}}]{2006astro.ph..9591A}
\bibinfo{author}{\bibfnamefont{A.}~\bibnamefont{{Albrecht}}},
  \bibinfo{author}{\bibfnamefont{G.}~\bibnamefont{{Bernstein}}},
  \bibinfo{author}{\bibfnamefont{R.}~\bibnamefont{{Cahn}}},
  \bibinfo{author}{\bibfnamefont{W.~L.} \bibnamefont{{Freedman}}},
  \bibinfo{author}{\bibfnamefont{J.}~\bibnamefont{{Hewitt}}},
  \bibinfo{author}{\bibfnamefont{W.}~\bibnamefont{{Hu}}},
  \bibinfo{author}{\bibfnamefont{J.}~\bibnamefont{{Huth}}},
  \bibinfo{author}{\bibfnamefont{M.}~\bibnamefont{{Kamionkowski}}},
  \bibinfo{author}{\bibfnamefont{E.~W.} \bibnamefont{{Kolb}}},
  \bibinfo{author}{\bibfnamefont{L.}~\bibnamefont{{Knox}}},
  \bibnamefont{et~al.}, \bibinfo{journal}{ArXiv Astrophysics e-prints}
  (\bibinfo{year}{2006}), \eprint{arXiv:astro-ph/0609591}.

\bibitem[{\citenamefont{{Komatsu} et~al.}(2011)\citenamefont{{Komatsu},
  {Smith}, {Dunkley}, {Bennett}, {Gold}, {Hinshaw}, {Jarosik}, {Larson},
  {Nolta}, {Page} et~al.}}]{2011ApJS..192...18K}
\bibinfo{author}{\bibfnamefont{E.}~\bibnamefont{{Komatsu}}},
  \bibinfo{author}{\bibfnamefont{K.~M.} \bibnamefont{{Smith}}},
  \bibinfo{author}{\bibfnamefont{J.}~\bibnamefont{{Dunkley}}},
  \bibinfo{author}{\bibfnamefont{C.~L.} \bibnamefont{{Bennett}}},
  \bibinfo{author}{\bibfnamefont{B.}~\bibnamefont{{Gold}}},
  \bibinfo{author}{\bibfnamefont{G.}~\bibnamefont{{Hinshaw}}},
  \bibinfo{author}{\bibfnamefont{N.}~\bibnamefont{{Jarosik}}},
  \bibinfo{author}{\bibfnamefont{D.}~\bibnamefont{{Larson}}},
  \bibinfo{author}{\bibfnamefont{M.~R.} \bibnamefont{{Nolta}}},
  \bibinfo{author}{\bibfnamefont{L.}~\bibnamefont{{Page}}},
  \bibnamefont{et~al.}, \bibinfo{journal}{\apjs}
  \textbf{\bibinfo{volume}{192}}, \bibinfo{eid}{18} (\bibinfo{year}{2011}),
  \eprint{1001.4538}.

\bibitem[{\citenamefont{{Pen} et~al.}(1998)\citenamefont{{Pen}, {Loeb}, and
  {Turok}}}]{1998ApJ...509..537P}
\bibinfo{author}{\bibfnamefont{U.-L.} \bibnamefont{{Pen}}},
  \bibinfo{author}{\bibfnamefont{A.}~\bibnamefont{{Loeb}}}, \bibnamefont{and}
  \bibinfo{author}{\bibfnamefont{N.}~\bibnamefont{{Turok}}},
  \bibinfo{journal}{\apj} \textbf{\bibinfo{volume}{509}}, \bibinfo{pages}{537}
  (\bibinfo{year}{1998}), \eprint{arXiv:astro-ph/9712178}.

\bibitem[{\citenamefont{{Espinosa} et~al.}(2010)\citenamefont{{Espinosa},
  {Konstandin}, {No}, and {Servant}}}]{2010JCAP...06..028E}
\bibinfo{author}{\bibfnamefont{J.~R.} \bibnamefont{{Espinosa}}},
  \bibinfo{author}{\bibfnamefont{T.}~\bibnamefont{{Konstandin}}},
  \bibinfo{author}{\bibfnamefont{J.~M.} \bibnamefont{{No}}}, \bibnamefont{and}
  \bibinfo{author}{\bibfnamefont{G.}~\bibnamefont{{Servant}}},
  \bibinfo{journal}{\jcap} \textbf{\bibinfo{volume}{6}}, \bibinfo{pages}{28}
  (\bibinfo{year}{2010}), \eprint{1004.4187}.

\bibitem[{\citenamefont{{Avelino} et~al.}(2002)\citenamefont{{Avelino},
  {Canavezes}, {de Carvalho}, and {Martins}}}]{2002APh....17..367A}
\bibinfo{author}{\bibfnamefont{P.~P.} \bibnamefont{{Avelino}}},
  \bibinfo{author}{\bibfnamefont{A.}~\bibnamefont{{Canavezes}}},
  \bibinfo{author}{\bibfnamefont{J.~P.~M.} \bibnamefont{{de Carvalho}}},
  \bibnamefont{and} \bibinfo{author}{\bibfnamefont{C.~J.~A.~P.}
  \bibnamefont{{Martins}}}, \bibinfo{journal}{Astroparticle Physics}
  \textbf{\bibinfo{volume}{17}}, \bibinfo{pages}{367} (\bibinfo{year}{2002}),
  \eprint{arXiv:astro-ph/0106245}.

\bibitem[{\citenamefont{{Maartens} et~al.}(1995)\citenamefont{{Maartens},
  {Ellis}, and {Stoeger}}}]{1995PhRvD..51.1525M}
\bibinfo{author}{\bibfnamefont{R.}~\bibnamefont{{Maartens}}},
  \bibinfo{author}{\bibfnamefont{G.~F.~R.} \bibnamefont{{Ellis}}},
  \bibnamefont{and} \bibinfo{author}{\bibfnamefont{W.~R.}
  \bibnamefont{{Stoeger}}}, \bibinfo{journal}{\prd}
  \textbf{\bibinfo{volume}{51}}, \bibinfo{pages}{1525} (\bibinfo{year}{1995}),
  \eprint{arXiv:astro-ph/9501016}.

\bibitem[{\citenamefont{{Zhang} and {Stebbins}}(2011)}]{2011PhRvL.107d1301Z}
\bibinfo{author}{\bibfnamefont{P.}~\bibnamefont{{Zhang}}} \bibnamefont{and}
  \bibinfo{author}{\bibfnamefont{A.}~\bibnamefont{{Stebbins}}},
  \bibinfo{journal}{Physical Review Letters} \textbf{\bibinfo{volume}{107}},
  \bibinfo{eid}{041301} (\bibinfo{year}{2011}), \eprint{1009.3967}.

\bibitem[{\citenamefont{{Shirokoff} et~al.}(2011)\citenamefont{{Shirokoff},
  {Reichardt}, {Shaw}, {Millea}, {Ade}, {Aird}, {Benson}, {Bleem}, {Carlstrom},
  {Chang} et~al.}}]{2011ApJ...736...61S}
\bibinfo{author}{\bibfnamefont{E.}~\bibnamefont{{Shirokoff}}},
  \bibinfo{author}{\bibfnamefont{C.~L.} \bibnamefont{{Reichardt}}},
  \bibinfo{author}{\bibfnamefont{L.}~\bibnamefont{{Shaw}}},
  \bibinfo{author}{\bibfnamefont{M.}~\bibnamefont{{Millea}}},
  \bibinfo{author}{\bibfnamefont{P.~A.~R.} \bibnamefont{{Ade}}},
  \bibinfo{author}{\bibfnamefont{K.~A.} \bibnamefont{{Aird}}},
  \bibinfo{author}{\bibfnamefont{B.~A.} \bibnamefont{{Benson}}},
  \bibinfo{author}{\bibfnamefont{L.~E.} \bibnamefont{{Bleem}}},
  \bibinfo{author}{\bibfnamefont{J.~E.} \bibnamefont{{Carlstrom}}},
  \bibinfo{author}{\bibfnamefont{C.~L.} \bibnamefont{{Chang}}},
  \bibnamefont{et~al.}, \bibinfo{journal}{\apj} \textbf{\bibinfo{volume}{736}},
  \bibinfo{eid}{61} (\bibinfo{year}{2011}), \eprint{1012.4788}.

\bibitem[{\citenamefont{{Reichardt} et~al.}(2011)\citenamefont{{Reichardt},
  {Shaw}, {Zahn}, {Aird}, {Benson}, {Bleem}, {Carlstrom}, {Chang}, {Cho},
  {Crawford} et~al.}}]{2011arXiv1111.0932R}
\bibinfo{author}{\bibfnamefont{C.~L.} \bibnamefont{{Reichardt}}},
  \bibinfo{author}{\bibfnamefont{L.}~\bibnamefont{{Shaw}}},
  \bibinfo{author}{\bibfnamefont{O.}~\bibnamefont{{Zahn}}},
  \bibinfo{author}{\bibfnamefont{K.~A.} \bibnamefont{{Aird}}},
  \bibinfo{author}{\bibfnamefont{B.~A.} \bibnamefont{{Benson}}},
  \bibinfo{author}{\bibfnamefont{L.~E.} \bibnamefont{{Bleem}}},
  \bibinfo{author}{\bibfnamefont{J.~E.} \bibnamefont{{Carlstrom}}},
  \bibinfo{author}{\bibfnamefont{C.~L.} \bibnamefont{{Chang}}},
  \bibinfo{author}{\bibfnamefont{H.~M.} \bibnamefont{{Cho}}},
  \bibinfo{author}{\bibfnamefont{T.~M.} \bibnamefont{{Crawford}}},
  \bibnamefont{et~al.}, \bibinfo{journal}{ArXiv e-prints}
  (\bibinfo{year}{2011}), \eprint{1111.0932}.

\bibitem[{\citenamefont{{Dunkley} et~al.}(2011)\citenamefont{{Dunkley},
  {Hlozek}, {Sievers}, {Acquaviva}, {Ade}, {Aguirre}, {Amiri}, {Appel},
  {Barrientos}, {Battistelli} et~al.}}]{2011ApJ...739...52D}
\bibinfo{author}{\bibfnamefont{J.}~\bibnamefont{{Dunkley}}},
  \bibinfo{author}{\bibfnamefont{R.}~\bibnamefont{{Hlozek}}},
  \bibinfo{author}{\bibfnamefont{J.}~\bibnamefont{{Sievers}}},
  \bibinfo{author}{\bibfnamefont{V.}~\bibnamefont{{Acquaviva}}},
  \bibinfo{author}{\bibfnamefont{P.~A.~R.} \bibnamefont{{Ade}}},
  \bibinfo{author}{\bibfnamefont{P.}~\bibnamefont{{Aguirre}}},
  \bibinfo{author}{\bibfnamefont{M.}~\bibnamefont{{Amiri}}},
  \bibinfo{author}{\bibfnamefont{J.~W.} \bibnamefont{{Appel}}},
  \bibinfo{author}{\bibfnamefont{L.~F.} \bibnamefont{{Barrientos}}},
  \bibinfo{author}{\bibfnamefont{E.~S.} \bibnamefont{{Battistelli}}},
  \bibnamefont{et~al.}, \bibinfo{journal}{\apj} \textbf{\bibinfo{volume}{739}},
  \bibinfo{eid}{52} (\bibinfo{year}{2011}), \eprint{1009.0866}.

\bibitem[{\citenamefont{{Zhang} et~al.}(2004)\citenamefont{{Zhang}, {Pen}, and
  {Trac}}}]{2004MNRAS.347.1224Z}
\bibinfo{author}{\bibfnamefont{P.}~\bibnamefont{{Zhang}}},
  \bibinfo{author}{\bibfnamefont{U.-L.} \bibnamefont{{Pen}}}, \bibnamefont{and}
  \bibinfo{author}{\bibfnamefont{H.}~\bibnamefont{{Trac}}},
  \bibinfo{journal}{\mnras} \textbf{\bibinfo{volume}{347}},
  \bibinfo{pages}{1224} (\bibinfo{year}{2004}),
  \eprint{arXiv:astro-ph/0304534}.

\bibitem[{\citenamefont{{Zahn} et~al.}(2011)\citenamefont{{Zahn}, {Reichardt},
  {Shaw}, {Lidz}, {Aird}, {Benson}, {Bleem}, {Carlstrom}, {Chang}, {Cho}
  et~al.}}]{2011arXiv1111.6386Z}
\bibinfo{author}{\bibfnamefont{O.}~\bibnamefont{{Zahn}}},
  \bibinfo{author}{\bibfnamefont{C.~L.} \bibnamefont{{Reichardt}}},
  \bibinfo{author}{\bibfnamefont{L.}~\bibnamefont{{Shaw}}},
  \bibinfo{author}{\bibfnamefont{A.}~\bibnamefont{{Lidz}}},
  \bibinfo{author}{\bibfnamefont{K.~A.} \bibnamefont{{Aird}}},
  \bibinfo{author}{\bibfnamefont{B.~A.} \bibnamefont{{Benson}}},
  \bibinfo{author}{\bibfnamefont{L.~E.} \bibnamefont{{Bleem}}},
  \bibinfo{author}{\bibfnamefont{J.~E.} \bibnamefont{{Carlstrom}}},
  \bibinfo{author}{\bibfnamefont{C.~L.} \bibnamefont{{Chang}}},
  \bibinfo{author}{\bibfnamefont{H.~M.} \bibnamefont{{Cho}}},
  \bibnamefont{et~al.}, \bibinfo{journal}{ArXiv e-prints}
  (\bibinfo{year}{2011}), \eprint{1111.6386}.

\bibitem[{\citenamefont{{Cabr{\'e}} et~al.}(2006)\citenamefont{{Cabr{\'e}},
  {Gazta{\~n}aga}, {Manera}, {Fosalba}, and {Castander}}}]{2006MNRAS.372L..23C}
\bibinfo{author}{\bibfnamefont{A.}~\bibnamefont{{Cabr{\'e}}}},
  \bibinfo{author}{\bibfnamefont{E.}~\bibnamefont{{Gazta{\~n}aga}}},
  \bibinfo{author}{\bibfnamefont{M.}~\bibnamefont{{Manera}}},
  \bibinfo{author}{\bibfnamefont{P.}~\bibnamefont{{Fosalba}}},
  \bibnamefont{and}
  \bibinfo{author}{\bibfnamefont{F.}~\bibnamefont{{Castander}}},
  \bibinfo{journal}{\mnras} \textbf{\bibinfo{volume}{372}},
  \bibinfo{pages}{L23} (\bibinfo{year}{2006}), \eprint{arXiv:astro-ph/0603690}.

\bibitem[{\citenamefont{{Zhang}}(2010)}]{2010MNRAS.407L..36Z}
\bibinfo{author}{\bibfnamefont{P.}~\bibnamefont{{Zhang}}},
  \bibinfo{journal}{\mnras} \textbf{\bibinfo{volume}{407}},
  \bibinfo{pages}{L36} (\bibinfo{year}{2010}), \eprint{1004.0990}.

\bibitem[{\citenamefont{{Ho} et~al.}(2008)\citenamefont{{Ho}, {Hirata},
  {Padmanabhan}, {Seljak}, and {Bahcall}}}]{2008PhRvD..78d3519H}
\bibinfo{author}{\bibfnamefont{S.}~\bibnamefont{{Ho}}},
  \bibinfo{author}{\bibfnamefont{C.}~\bibnamefont{{Hirata}}},
  \bibinfo{author}{\bibfnamefont{N.}~\bibnamefont{{Padmanabhan}}},
  \bibinfo{author}{\bibfnamefont{U.}~\bibnamefont{{Seljak}}}, \bibnamefont{and}
  \bibinfo{author}{\bibfnamefont{N.}~\bibnamefont{{Bahcall}}},
  \bibinfo{journal}{\prd} \textbf{\bibinfo{volume}{78}}, \bibinfo{eid}{043519}
  (\bibinfo{year}{2008}), \eprint{0801.0642}.

\end{thebibliography}

\label{lastpage}

\end{document}